\documentstyle[epsfig]{mn2e}
\begin{document}
\input BoxedEPS.tex
\newcommand{\aip}{{\small ${\cal AIPS}$}}
\newcommand{\gtsim}{\mbox{{\raisebox{-0.4ex}{$\stackrel{>}{{\scriptstyle\sim}}
$}}}}
\newcommand{\ltsim}{\mbox{{\raisebox{-0.4ex}{$\stackrel{<}{{\scriptstyle\sim}}
$}}}}
\newcommand{\s}{$\stackrel{\rm s}{.}$}
\newcommand{\h}{$^{\rm h}$}
\newcommand{\m}{$^{\rm m}$}
\newcommand{\pp}{$\stackrel{\prime\prime}{.}$}
\newcommand{\de}{$^{\circ}$}
\newcommand{\p}{$^{\prime}$}
\newcommand{\arc}{$^{\prime\prime}$}
\newcommand{\marc}{^{\prime\prime}}
\newcommand{\rs}{{\em $r_s$}}
\newcommand{\DPM}{{\em DPM}}
\newcommand{\alf}{{\displaystyle\biggl({\nu_{\rm h} \over \nu_{\rm l}}\biggr)^{\alpha}} }

\newcommand{\figstart}[1]
    { \begin{figure}[htb]
      \begin{picture}(0,#1) }
\newcommand{\figend}[4]
    { \end{picture}
      \special{#1}
      \caption[#2]{#3}
      \label{#4}
      \end{figure} }
\newcommand{\fig}[5]
    { \figstart{#1}
      \figend{#2}{#3}{#4}{#5} }
\newcommand{\bHS}{\beta_{\mbox{\scriptsize HS}}}
\newcommand{\bBF}{\beta_{\mbox{\scriptsize BF}}}
\newcommand{\nT}{\nu_{\mbox{\scriptsize T}}}
\newcommand{\et}{E_{\mbox{\scriptsize T}}}
\newcommand{\nTn}{\nu_{\mbox{\scriptsize Tn}}}
\newcommand{\nTf}{\nu_{\mbox{\scriptsize Tf}}}
\newcommand{\tn}{\tau_{x\mbox{\scriptsize n}}}
\newcommand{\tf}{\tau_{x\mbox{\scriptsize f}}}
\newcommand{\xn}{x_{\mbox{\scriptsize n}}}
\newcommand{\xf}{x_{\mbox{\scriptsize f}}}
\newcommand{\yn}{y_{\mbox{\scriptsize n}}}
\newcommand{\yf}{y_{\mbox{\scriptsize f}}}
\newcommand{\lln}{l_{\mbox{\scriptsize n}}}
\newcommand{\llf}{l_{\mbox{\scriptsize f}}}
\newcommand{\Dn}{f(\Delta_{\mbox{\scriptsize n}})}
\newcommand{\Df}{f(\Delta_{\mbox{\scriptsize f}})}
\newcommand{\B}{\mbox{$B$}}
\newcommand{\Bo}{\mbox{$B$}_{0}}

\SetEPSFDirectory{/scratch/sbgs/figures/hst/}
\SetRokickiEPSFSpecial
\HideDisplacementBoxes

\title[Hyperluminous infrared galaxies from IIFSCz]{Hyperluminous infrared galaxies from IIFSCz}
\author[Rowan-Robinson M.]{Michael Rowan-Robinson and Lingyu Wang\\
Astrophysics Group, Blackett Laboratory, Imperial College of Science 
Technology and Medicine, Prince Consort Road,\\ 
London SW7 2AZ\\
Astronomy Centre, Department of Physics and Astronomy, University of Sussex, Falmer, Brighton, BN1 9QH}
\maketitle
\begin{abstract}
We present a catalogue of 179 hyperluminous infrared galaxies (HLIRGs) from the Imperial IRAS-FSS 
Redshift (IIFSCz) Catalogue.  Of the 92 with detections in at least two far infrared bands, 62 are dominated by an
M82-like starburst, 22 by an Arp220-like starburst and 8 by an AGN dust torus. On the basis of previous
gravitational lensing studies and an examination of HST archive images for a further 5 objects, we
estimate the fraction of HLIRGs that are significantly lensed to be 10-30$\%$.

We show simple infrared template fits to the SEDs of 23 HLIRGs with 
spectroscopic redshifts and at least 5 photometric bands.  Most can be fitted with a combination 
of two simple templates: an AGN dust torus and an M82-like starburst.  In the optical, 17 of the objects
are fitted with QSO templates, several with quite strong extinction. There are 5 objects fitted with 
galaxy templates in the optical, two of which show evidence for AGN dust tori and so presumably contain
Type 2 (edge-on) QSOs. The remaining object is fitted with a galaxy template in the optical, but 
is of such high luminosity that this classification would be plausible only if there were very 
strong lensing. 20 of the 23 objects (87$\%$) show evidence of an AGN either from the optical continuum
or from the signature of an AGN dust torus, but the starburst component is the dominant contribution to
bolometric luminosity in 14 out of 23 objects (61 $\%$).  The implied star-formation rates, even after correcting 
for lensing magnification, are in excess of 1000 $M_{\odot} yr^{-1}$.

We use infrared
template-fitting models  to predict fluxes for all HLIRGs at submillimetre wavelengths, and show
predictions at 350 and 850 $\mu$m.  Most would have 850 $\mu$m fluxes brighter than 5 mJy so should 
be easily detectable with current submillimetre telescopes.
At least 15 $\%$ should be detectable in the {\it Planck} all-sky survey at 350 $\mu$m and all {\it Planck} 
all-sky survey sources with z $<$ 0.9 should be IIFSCz sources.

From the luminosity-volume test we find that HLIRGs show strong evolution.  A simple exponential luminosity 
evolution applied to all HLIRGs would be consistent with
the luminosity functions found in redshift bins 0.3-0.5, 0.5-1 and 1-2.  The evolution-corrected luminosity
function flattens towards higher luminosities perhaps indicating a different physical mechanism is at work
compared to lower luminosity starbursts.  In principle this could be gravitational lensing though previous 
searches with HST have, perhaps surprisingly, not shown lensing to be widely prevalent in HLIRGs. 

\end{abstract}


\section{Introduction}

Rowan-Robinson (2000, hereafter RR2000) defined hyperluminous infrared galaxies to be those with
bolometric infrared luminosities $L_{ir} > 10 ^{13} L_{\odot}$.  He discussed a sample
of 39 such galaxies found either by follow-up of the IRAS survey or through submillimetre
observations, and modelled the infrared and submillimetre spectral energy distributions (SEDs) for
those that were well-observed.  Submillimetre observations and more detailed SED models for 13 
hyperluminous infrared galaxies (HLIRGs) were reported by Farrah et al (2002a) and ISO data for 4 HLIRGs was
discussed by Verma et al (2002).  A more detailed SED model for the prototypical HLIRG IRAS F10214+4724
was give by Efstathiou (2006).  Recently Ruiz et al (2010) have modelled the SEDs of 13 hyperluminous 
galaxies which are X-ray sources.
Many of these galaxies show evidence of AGN dust torus emission at rest-frame 
wavelengths 3-30 $\mu$m, but the submillimetre emission is almost always due to star-formation,
an interpretation that is supported by the large gas-masses found in these galaxies (see RR2000
for a summary).  7 of the 39 objects in RR2000 were already known to be gravitationally lensed.
A further 9 hyperluminous infrared galaxies were imaged with HST by Farrah et al (2002b), but
none were found to be lensed. 
Even after correction for magnification by gravitational lensing, most of these galaxies imply star 
formation rates $> 1000 M_{\odot} yr^{-1}$.

Since many of the hyperluminous infrared galaxies in RR2000 were discovered through their
submillimetre radiation, it is clearly of interest to consider whether all or most submillimetre
galaxies are hyperluminous.  Many discussions of the nature of 'submillimetre galaxies' assume the latter
to be the case. However there is growing evidence that the submillimetre population is quite diverse.
In their models for the SEDs of a complete sample of 850 $\mu$m sources from the SHADES
survey, Clements et al (2008) find a diverse range of SED types, including M82 starbursts, 'cirrus'
(ie quiescent) galaxies, and AGN dust tori, as well as Arp220-type starbursts. Efstathiou and
Rowan-Robinson (2003) have pointed out that the SEDs of as many as half of submillimetre galaxies
could be interpreted in terms of a cirrus model.  In modelling source-counts from 8-1100 $\mu$m,
and the far infrared and submillimetre background radiation, Rowan-Robinson (2008) concludes that 
at S(850) = 8 mJy there are approximately equal numbers of M82 and Arp220 starbursts.  However
cirrus galaxies, in these models, dominate the counts only at S(850) $>$ 100 mJy and at S(850) $<$ 1 mJy.

The creation of the Imperial IRAS FSC Redshift Catalogue (IIFSCz, Wang and 
Rowan-Robinson 2009a) gives us the opportunity
to define an almost all-sky ($|b| > 20^0$) catalogue of hyperluminous infrared galaxies,
sampled at 60 $\mu$m.  This allows us to study the evolution and luminosity function of this
population and to predict what will be seen in submillimetre surveys to be undertaken by
{\it Planck} and {\it Herschel}.

A spatially flat cosmological model with $\Lambda$ = 0.7, $h_0$=0.72 has been used throughout.

\section{The Catalogue}

Our catalogue of hyperluminous IRAS galaxies, with $L_{ir} > 10^{13} L_{\odot}$,
derived from IIFSCz, contains 179 galaxies.  The catalogue, together with the full IIFSCz
Catalogue, is available at http:$//$astro.imperial.ac.uk$/ \sim$mrr$/$fss$/$hyper.

\begin{figure*}
\epsfig{file=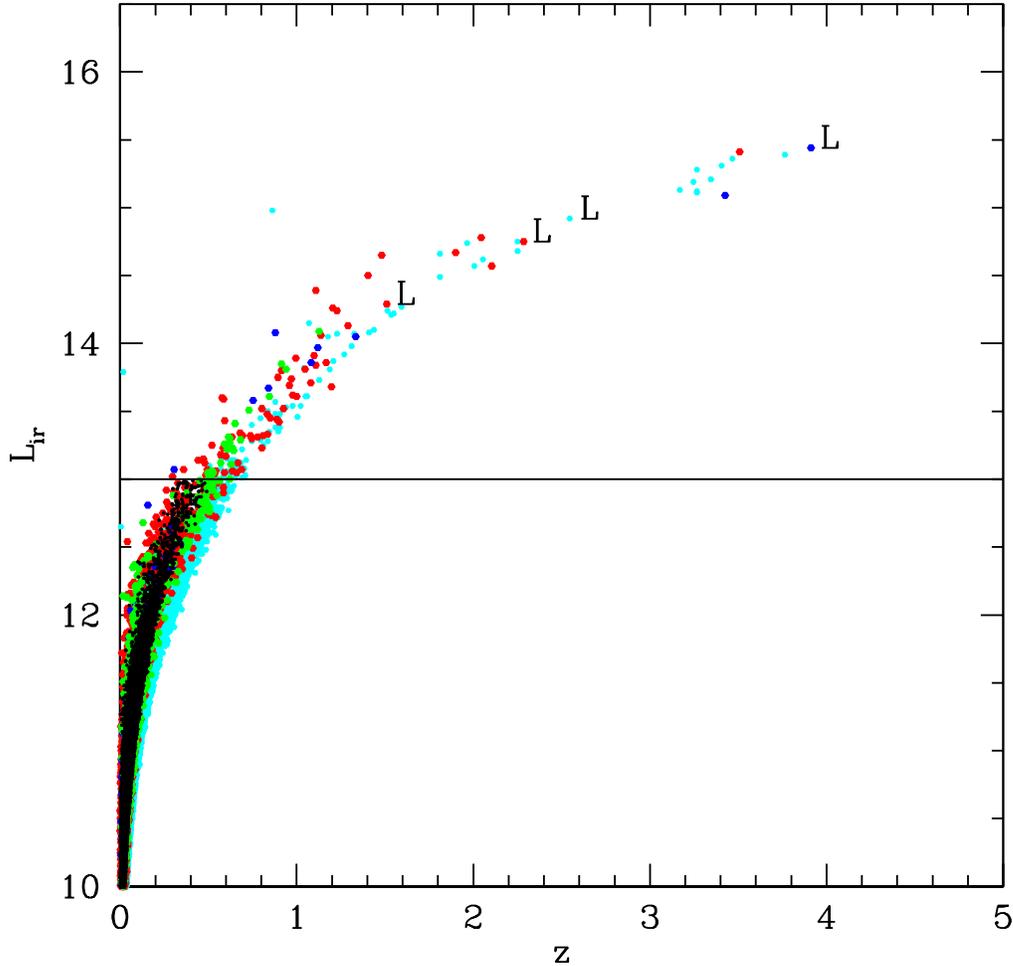,angle=0,width=14cm}
\caption{Bolometric infrared luminosity versus redshift for IIFSCz galaxies.
  Colour coding indicates dominant contribution to bolometric infrared luminosity.
Black: cirrus (quiescent); red: M82 starburst; green: Arp220
starburst; blue: AGN dust torus; cyan: M82 starburst fitted to 60 $\mu$m only.  Known
gravitationally lensed systems are labelled L.
}
\end{figure*}

Of the 25 hyperluminous galaxies in Tables 1 and 2 of RR2000, found from 60 or 850 $\mu$m surveys, 
5 were too weak at 60 $\mu$m 
to be detected in the FSS, and 2 lie at $|b| < 20^0$.  Of the remaining 18, 14 are in the present 
catalogue, 3 lie just below the $10^{13} L_{\odot}$
limit, and one has had its redshift revised to a lower value.
On the basis of the 12 hyperluminous galaxies detected from unbiased IRAS surveys, Rowan-Robinson (2000)
estimated that there would be 100-200 such galaxies with S(60) $>$ 200 mJy over the whole sky .   

Figure 1 shows a plot of $L_{ir}$ versus redshift for IIFSCz galaxies, colour-coded by the dominant
infrared template type. Note that in fitting infrared templates we 
specifically exclude the possibility of an HLIRG having a cirrus template.
Of the 92 galaxies with detections in at least two infrared bands, the infrared luminosities of 62 are dominated 
by an M82-like starburst, 22 by an Arp 220-like starburst, and 8 by an AGN dust torus.  Where only one
infrared band is available we have estimated the luminosity using an M82 template.

We have spectroscopic redshifts from the literature for 58 (32$\%$) HLIRGs, photometric redshifts for 
the remainder.  We have 
spectroscopic redshifts for 18 out of 46 (39$\%$) of galaxies with infrared luminosities, $L_{ir} > 10^{14} L_{\odot}$. 
Four  hyperluminous infrared galaxies from this catalogue are known to be gravitationally lensed:
F08105+2554=HS0810+2554 (Reimers et al 2002), F08279+5255 (Downes et al 1998), F10214+4724 
(Eisenhardt et al 1996) and F14132+1144=H1413+117 (Yun et al 1997).  The first of these does not have a lens model, 
but the other three remain in the hyperluminous category when corrected for their estimated far infrared lensing 
magnifications of 14, 10 and 10, respectively (see Table 1).  We have adopted a far infrared lensing magnification
of 10 for F08105+2554 by analogy with the very similar system PG1115+080 (Reimers 2002).
A further six hyperluminous galaxies were found by Farrah et al (2002)
not to be lensed (F00235+1024, F10026+4949, F12509+3122, F14218+3845, F15307+3252, 
F16348+7037=PG1634+706).  
We examined the HST archive images for a further five hyperluminous galaxies ( F01175-2025,
F09105+4108, F14165+0642, F09105+4108, F18216+6419) and found no evidence of lensing,
though F09105+4108 and F09105+4108 have highly disturbed images indicating a major merger.
Thus in the sample for which we have information on lensing, the fraction which are significantly lensed is
4/15 (27$\%$).
On the basis of the relatively unbiassed search for lensing in
HLIRGs by Farrah et al (2002), the fraction of HLIRGs which are significantly lensed may be no
greater than 10$\%$.  However this is clearly a very uncertain estimate and it would be worthwhile searching 
for new lenses in this sample, especially amongst the galaxies with infrared luminosities, $L_{ir} > 10^{14} L_{\odot}$.
The lensing rate might be expected to be higher for high redshift (z$>$2) galaxies.  There are 20 such galaxies in
our catalogue and 3 are already known to be lensed.

\begin{figure*}
\epsfig{file=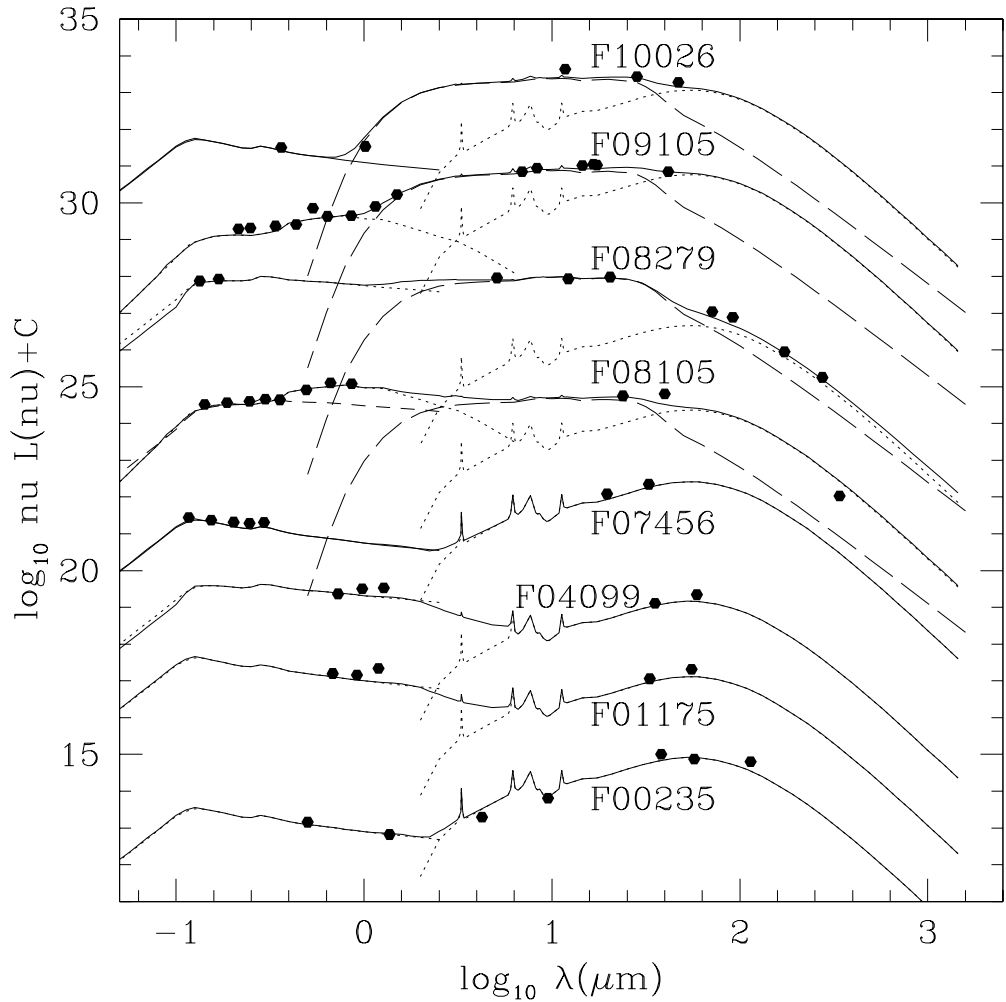,angle=0,width=14cm}
\caption{SEDs for 8 hyperluminous galaxies with spectroscopic redshifts. The template fits used in the infrared are  
M82 starburst (dotted curve), AGN dust torus (long-dashed curve).  Parameters for fits given in Table 1.
}
\end{figure*}

\begin{figure*}
\epsfig{file=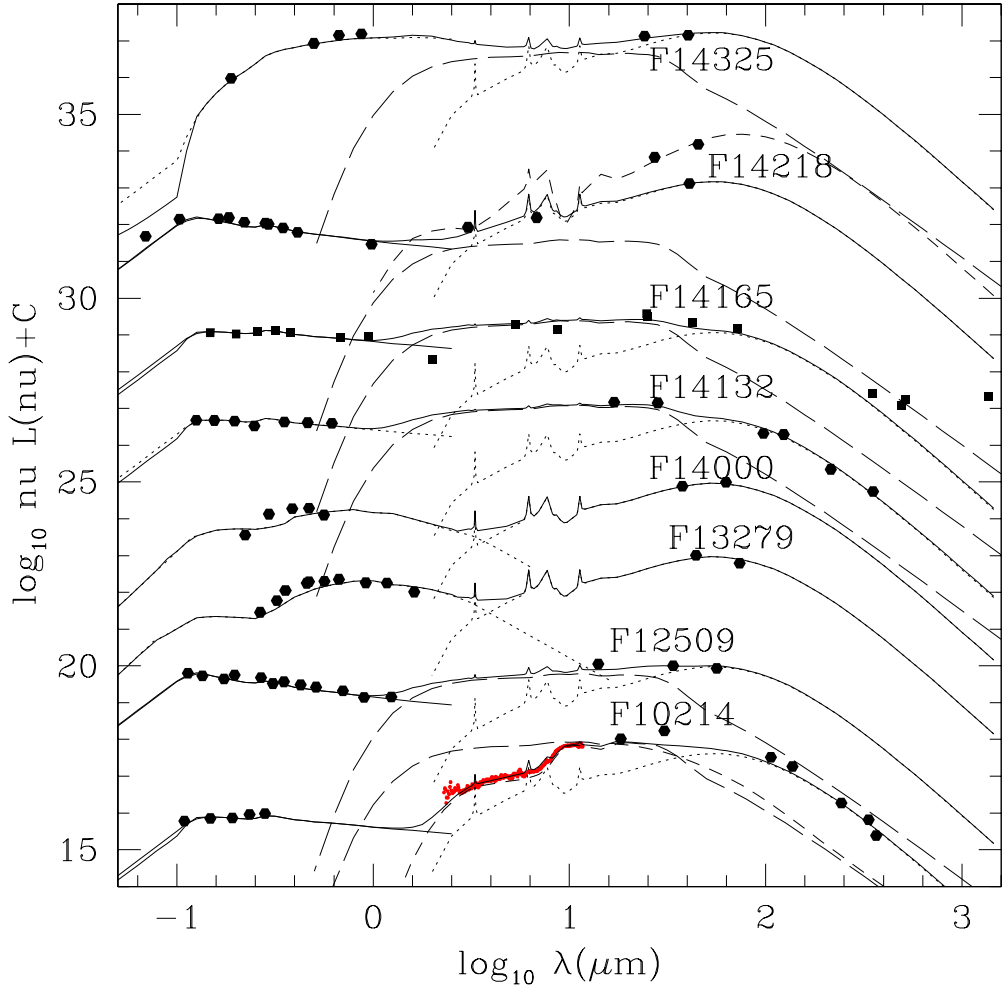,angle=0,width=14cm}
\caption{SEDs for 8 hyperluminous galaxies. Parameters for fits given in Table 1.  The IRS data
for F10214+4724 are indicated in red and the Efstathiou (2006) torus model for this object is shown as 
a short-dashed locus.
}
\end{figure*}

\begin{figure*}
\epsfig{file=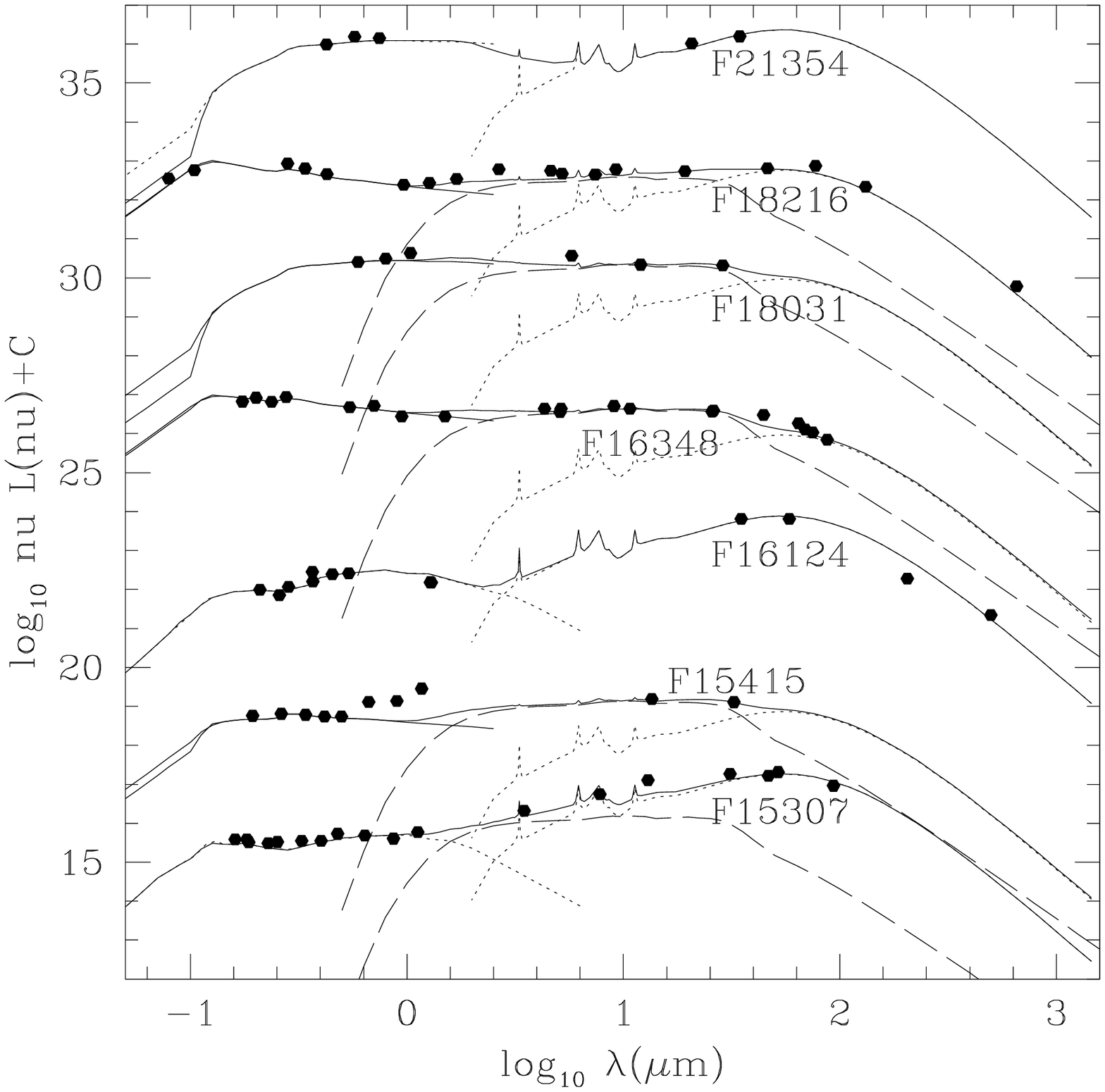,angle=0,width=14cm}
\caption{SEDs for 7 hyperluminous galaxies. Parameters for fits given in Table 1.  
}
\end{figure*}

\section{Spectral energy distributions and predicted submillimetre fluxes}

Detailed modelling of SEDs of hyperluminous galaxies was presented by Rowan-Robinson (2000)
and Farrah et al (2002b), and the latter presented some new submillimetre observations for 13
hyperluminous galaxies.  Efstathiou (2006) has modelled the prototypical hyperluminous galaxy 
F10214+4724, for which Spitzer IRS data was obtained by Teplitz et al (2006).  Figures 2-4
show SEDs for 23 hyperluminous galaxies, which have spectroscopic redshifts and  
photometric data in at least 5 bands, including at least two infrared bands.  Data have been 
compiled from the literature.  Table 1 gives physical
parameters for the 23 sources.  The columns are: name, redshift, lensing magnification, 
luminosities in starburst, dust torus, and 
optical galaxy/QSO components, optical template type, visual extinction, star-formation rate.  
Luminosities in this table have been corrected for the estimated lensing magnification.
12 have strong AGN dust tori, while 9 show no evidence for a dust torus, though this may be partly because
of a lack of observational data at rest-frame wavelengths 3-30 $\mu$m .  In fact the infrared SEDs 
of all 23 objects can be fitted with mixtures of just two components, an AGN dust torus and an M82 starburst. 
The model fits in Figs 2-4 broadly agree with the automatic template fits just to IRAS data: the relative
luminosities of starburst and dust torus are slightly modified in some cases when fitting to the
additional far infrared and submillimetre photometry here.  No galaxies with Arp 220
fits in the IIFSCz Catalogue had a sufficient number of photometric bands to be plotted in these figures.

In the optical, 17 of the objects
are fitted with QSO templates, three with strong extinction ($A_V > 0.5$). There are 5 objects fitted with 
galaxy templates in the optical, two of which show evidence for AGN dust tori and so presumably contain
Type 2 (edge-on) QSOs. The remaining object (F08105+2554) is best fitted with a galaxy template in the optical, but 
is of such high optical luminosity that this classification would be plausible only if there were very 
strong lensing.  As this is a known quadruple lensed system (Reimers et al 2002), this is a possibility
(our estimate of the lensing magnification is 10 - see section 2): a QSO fit is 
also shown, which fits the optical data but not the JHK data.  For F14218+3845 there is a discrepancy 
between the IRAS fluxes and the 90 $\mu$m flux measured with ISO.  Our preferred model is the more
conservative fit to the ISO data.
20 of the 23 objects (87$\%$) show evidence of an AGN either from the optical continuum
or from the signature of an AGN dust torus, but the starburst component is the dominant contribution to
bolometric luminosity in 14 out of 23 objects (61 $\%$).  Even if the QSO is not the dominant source of the 
extreme infrared luminosities, there is clearly a strong case for a link between the very high rates
of star-formation implied by these galaxies and the growth of massive black holes
(cf Sanders et al 1988, Kauffmann and Haehnelt 2000, Vielleux et al 2009).  However it should be borne 
in mind that restricting the sample of Fig 2 and Table 1 to objects with spectroscopic redshifts inevitably
biases the sample towards QSOs. 

\begin{figure*}
\epsfig{file=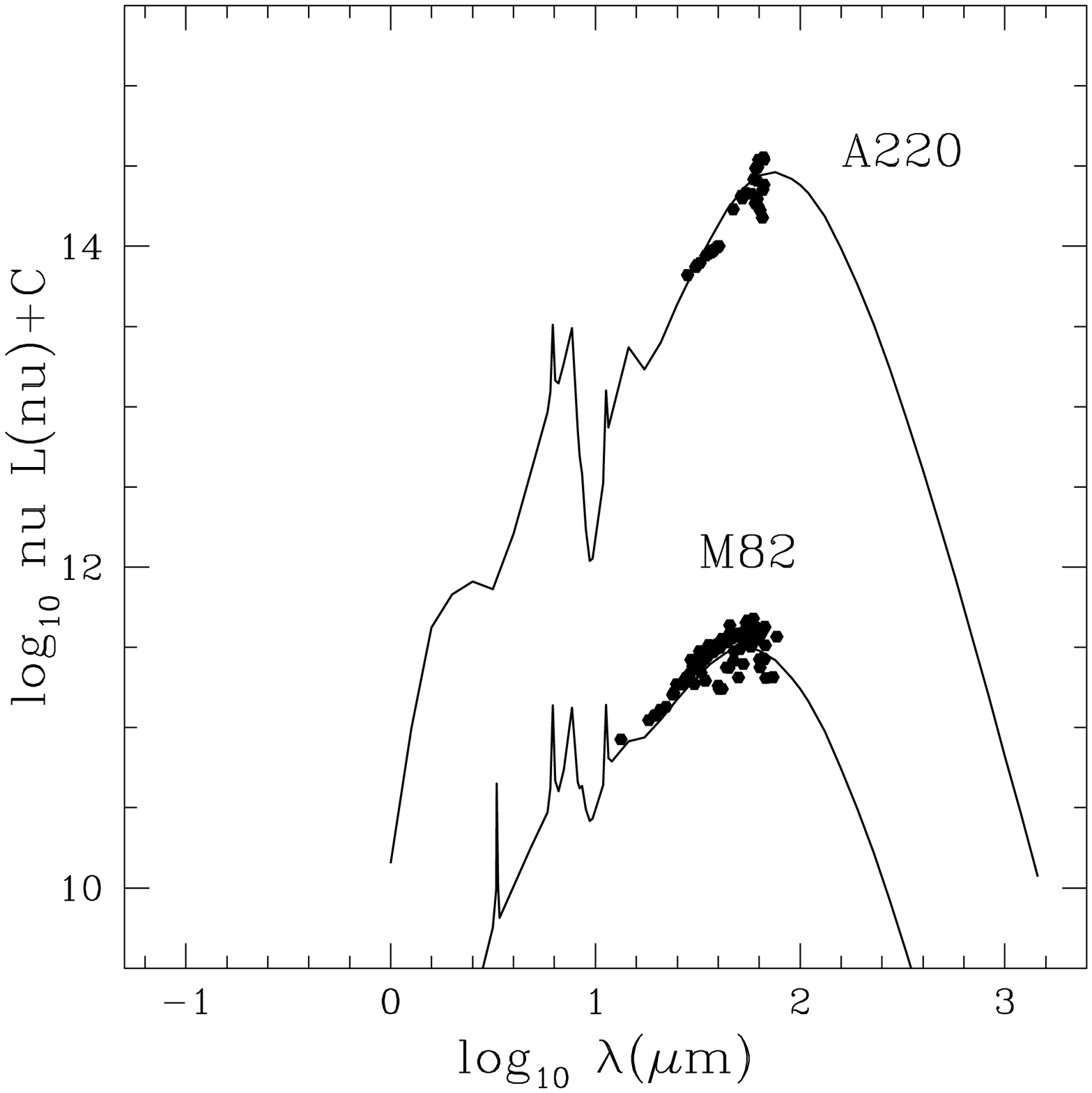,angle=0,width=14cm}
\caption{Composite SEDs for hyperluminous galaxies, detected in at least two far infrared bands, which
select either M82 or Arp 220 starburst templates.  
}
\end{figure*}

Our simple dust torus template does not fit
the data for F10214+4724, which Efstathiou (2006) suggests is a highly edge-on system.  We have used the
Efstathiou (2006) edge-on dust-torus model in our fit to F10214+4724 (Fig 3).  The J,H,K data
for F15415+1633 are brighter than our model predicts, and this is true also for F08105+2554 if a QSO model
is preferred for the optical SED.   Three strong radio sources
which have very flat SEDs and are strongly variable, OJ287, 3C345 and 3C446, have
not been included in the SED plots.

For QSOs the ratio of torus luminosity to optical luminosity, $L_{tor}/L_{opt}$, can be interpreted as the dust covering factor
(Rowan-Robinson et al 2009).  In this sample there are several objects with QSO-like optical SEDs for which $L_{tor}$ is significantly 
greater than $L_{opt}$ (F10026+4949, F10214+4724, F15307+3253) and for these we must assume that we are seeing only
a small fraction of the QSO optical light, probably via scattering off the dust torus (cf Polletta et al 2006).
 
\begin{table*}
\caption{Parameters for SED fits}
\begin{tabular}{lllllllllll}
object name  & z & lensing & $lg L_{sb}$ & $lg L_{tor}$ & $lg L_{opt}$ & type & $A_V$ & lg sfr\\
&&magn.&&&&&&\\
&&&&&&&&\\
F00235+1024 & 0.575 & 1 & 13.01 & & 11.61 & QSO &  & 3.31 \\
F01175-2025 & 0.810 & 1 &13.25 & & 13.75 & QSO & & 3.55 \\
F04099-7514 & 0.694 & ? & 13.28 & & 14.08 & QSO & 0.20 & 3.58 \\
F07456+4736 & 2.045 & ? & 14.58 & & 13.53 & QSO & & 4.88 \\
F08105+2554 & 1.510 & 10? & 12.57 & 12.97 & 13.27 & Scd & & 3.87 \\
 & & & 12.57 & 12.97 & 13.37 & QSO & 0.4 & 3.87 \\
F08279+5255* & 3.912 & 14 & 12.79 & 14.19 & 14.54 & QSO & 0.3 & 4.09 \\
F09105+4108 & 0.442 & 1? & 12.83 & 13.03 & 11.73 & Scd & & 3.13 \\
F10026+4949 & 1.120 & 1 & 13.23 & 13.63 & 11.78 & QSO & & 3.53  \\
&&&&&&&&\\
F10214+4724* & 2.286 & 10 & 12.85 & 13.04 & 11.50 & QSO & 0.2 & 4.15 \\
F12509+3122 & 0.780 & 1 & 13.04 & 11.99 & 12.89 & QSO & & 3.34 \\
F13279+3402 & 0.360 & ? & 13.01 & & 12.31 & Sab & & 3.31 \\
F14000+0158 & 0.591 & ? & 13.06 & & 12.36 & E & & 3.36 \\
F14132+1144* & 2.546 & 10 & 12.91 & 13.41 & 13.31 & QSO & 0.2 & 4.21 \\
F14165+0642 & 1.437 & 1 & 13.41 & 13.56 & 13.66 & QSO & 0.2 & 3.56\\
F14218+3845 & 1.205 & 1 &  13.34 & 11.84 & 12.44 & QSO & & 3.64 \\
 & & & [14.54(A220) & & 12.44 & QSO & & 4.84]\\
F14325-0447 & 1.482 & ? & 14.42 & 13.97 & 15.12 & QSO & & 4.72\\
&&&&&&&&\\
F15307+3252 & 0.926 & 1 & 13.41 & 12.41 & 12.06 & sb & 0.1 & 3.71 \\
F15415+1633 & 0.850 & ? & 13.01 & 13.36 & 13.41 & QSO & 0.35 & 3.31 \\
F16124+3241 & 0.710 & 1? & 13.00 & & 11.63 & Scd & & 3.30 \\
F16348+7037 & 1.334 & 1 & 13.16 & 13.91 & 14.36 & QSO & 0.1 & 3.46 \\
F18031+7517 & 1.083 & ? & 13.13 & 13.48 & 14.28 & QSO & 1.0 & 3.43 \\
F18216+6419 & 0.297 & 1 & 12.79 & 12.69 & 12.99 & QSO & & 3.09\\
F21354-1432 & 1.900 & ? & 14.59 &  & 14.99 & QSO & 1.0 & 4.89\\
\end{tabular}
\end{table*}

To test the validity of our template fits to the IRAS data we show in Fig 5 composite spectra of HLIRGs,
detected in at least two far infrared bands, which select M82 or Arp 220 templates.  The distinction 
between these two SED types appears to be genuine.

Wang and Rowan-Robinson (2009a) fitted four infrared 
templates to the far infrared SEDs of the whole IIFSCz Catalogue and used these to
predict submillimetre fluxes at a range of wavelengths and we have used these template
fits to predict 350 and 850 $\mu$m fluxes, and to investigate evolution.  Efstathiou and Rowan-Robinson (2009)
have analyzed Spitzer-IRS mid-infrared spectroscopic data for infrared galaxies, in particular the diagnostic diagram
of Spoon et al (2006), in terms of sequences of starburst, cirrus (quiescent), and AGN dust tori models.  They conclude 
that infrared template fitting could be improved by including 
additional templates corresponding to very young and very old starbursts.  However the 
shortage of far infrared photometric data (rarely more than 2 detected bands) for the 
present sample means this would not be very meaningful at this stage.



We have used our 4-template fits to predict submillimetre fluxes for the IIFSCz catalogue.
Figure 6 shows predicted 850 $\mu$m flux versus z, with hyperluminous galaxies indicated by open squares.
Most of the hyperluminous galaxies in our catalogue have predicted 850 $\mu$m fluxes brighter than 
5 mJy and should be easily detectable with current submillimetre telescopes.

Figure 7 shows predicted 350 mu flux versus z.  We have indicated the estimated PLANCK
all-sky survey detection limit.  Out to z $\sim$0.9 all sources detected by Planck would be in
the IIFSCz (for $|b| > 20^o$).  At least 15 $\%$ of hyperluminous infrared galaxies in
our catalogue should be detectable by PLANCK.  Clearly many other luminous
infrared galaxies at z $>$ 1.0 will also be detected by PLANCK.

Individual flux predictions are very uncertain, since they are based on 25-100 $\mu$m fluxes
and in many cases on 60 $\mu$m detections only.  For example if an M82 starburst is incorrectly
classified as an Arp220 type, the submillimetre fluxes and bolometric luminosities can
be overestimated by an order of magnitude.

\begin{figure*}
\epsfig{file=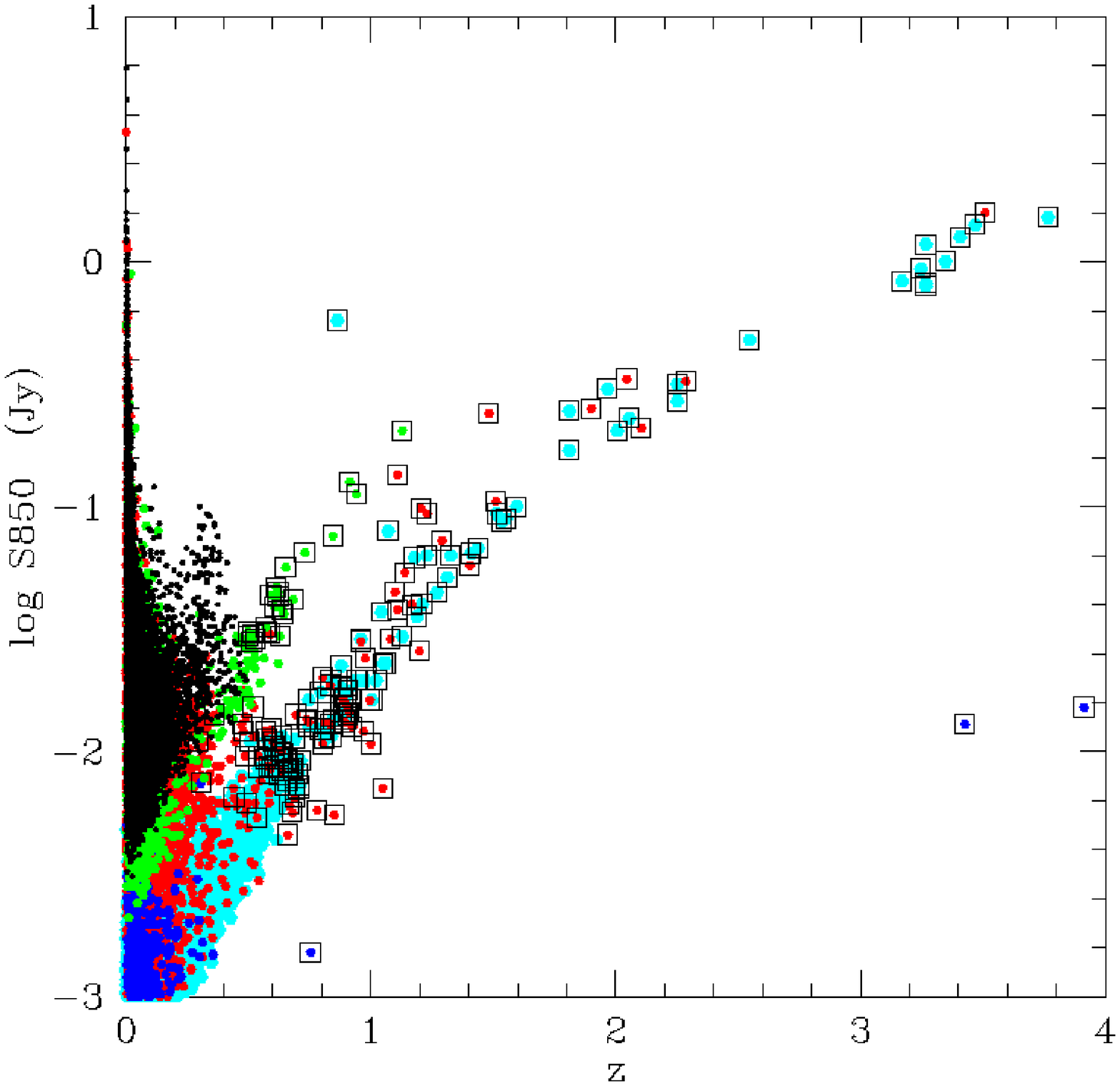,angle=0,width=14cm}
\caption{Predicted 850 $\mu$m flux versus redshift for IIFSCz galaxies, with hyperluminous
galaxies shown as open squares.  Black: cirrus (quiescent); red: M82 starburst; green: Arp220
starburst; blue: AGN dust torus; cyan: M82 starburst fitted to 60 $\mu$m only.
}
\end{figure*}

\begin{figure*}
\epsfig{file=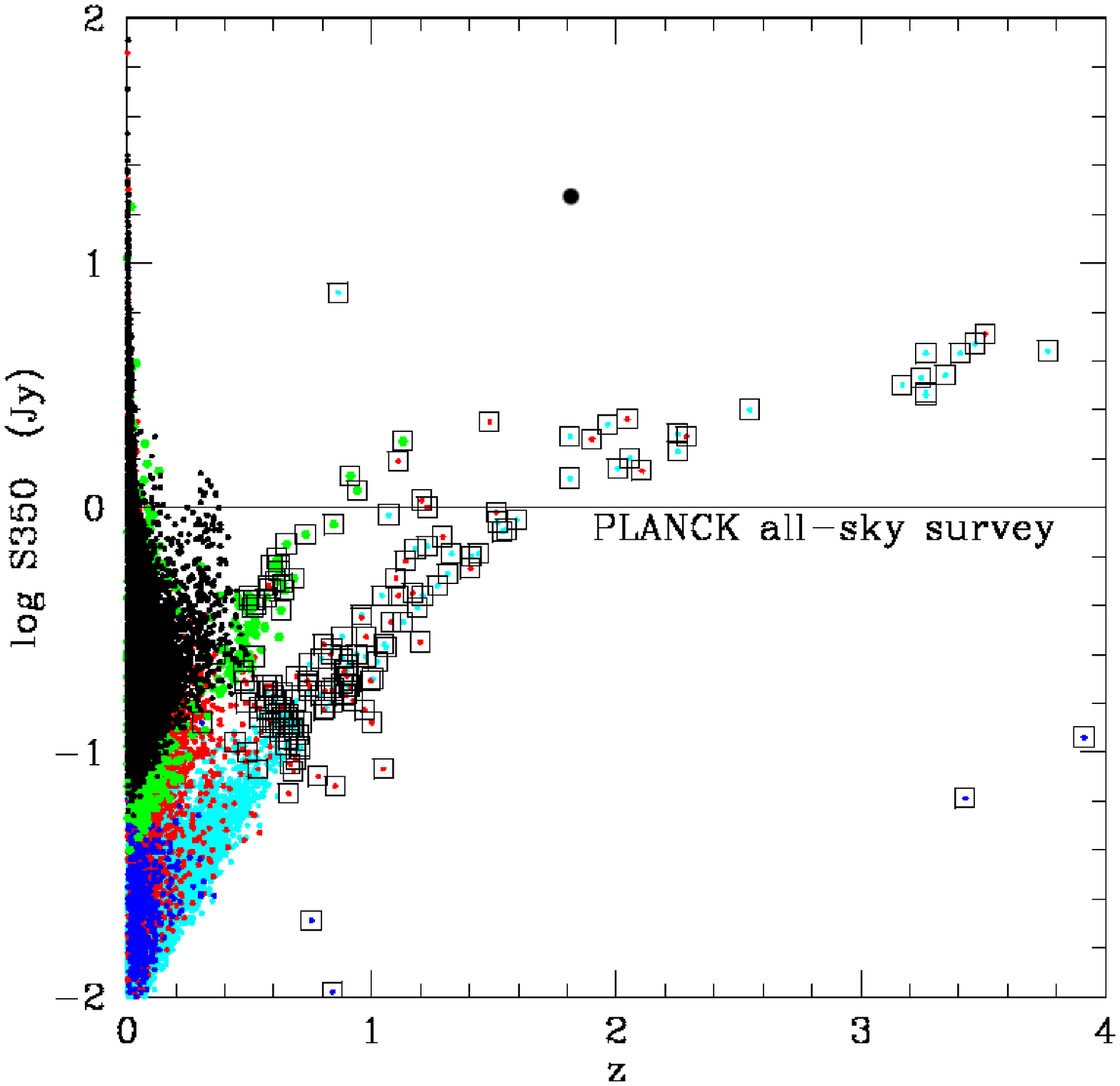,angle=0,width=14cm}
\caption{Predicted 350 $\mu$m flux versus redshift for IIFSCz galaxies, with hyperluminous
galaxies shown as open squares.  Key for coloured symbols as in Fig 2.   The horizontal line
shows the Planck Blue Book (2005) estimate of the sensitivity of the all-sky 350 $\mu$m 
survey (10-$\sigma$).
}
\end{figure*}

\section{Evolution and luminosity functions}

To investigate whether there is evidence for evolution in luminous infrared galaxies we
first plot infrared luminosity versus comoving volume, Fig 8L.
If no evolution was present then in each band of luminosity there would be a uniform
distribution of objects with volume (Rowan-Robinson 1968).  Clearly for ULIRGs and HLIRGs there is strong evolution.
Fig 8R shows the corresponding plot corrected for luminosity evolution with Q = 3.0.
 The solid curves denote the FSS 90$\%$ completeness limit of S60 = 0.36 Jy
and only sources brighter than this limit have been included in these plots.
\begin{figure*}
\epsfig{file=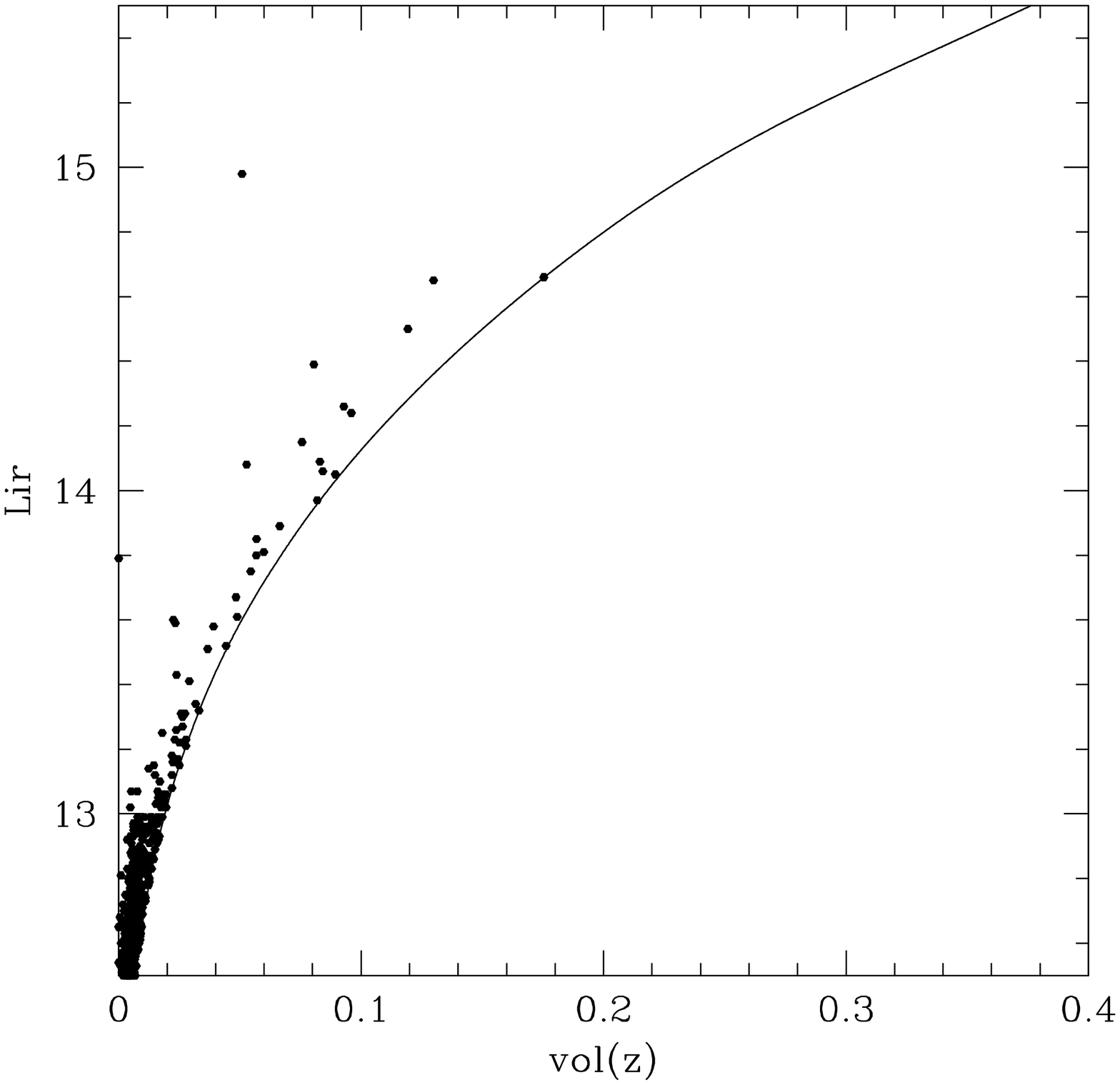,angle=0,width=7cm}
\epsfig{file=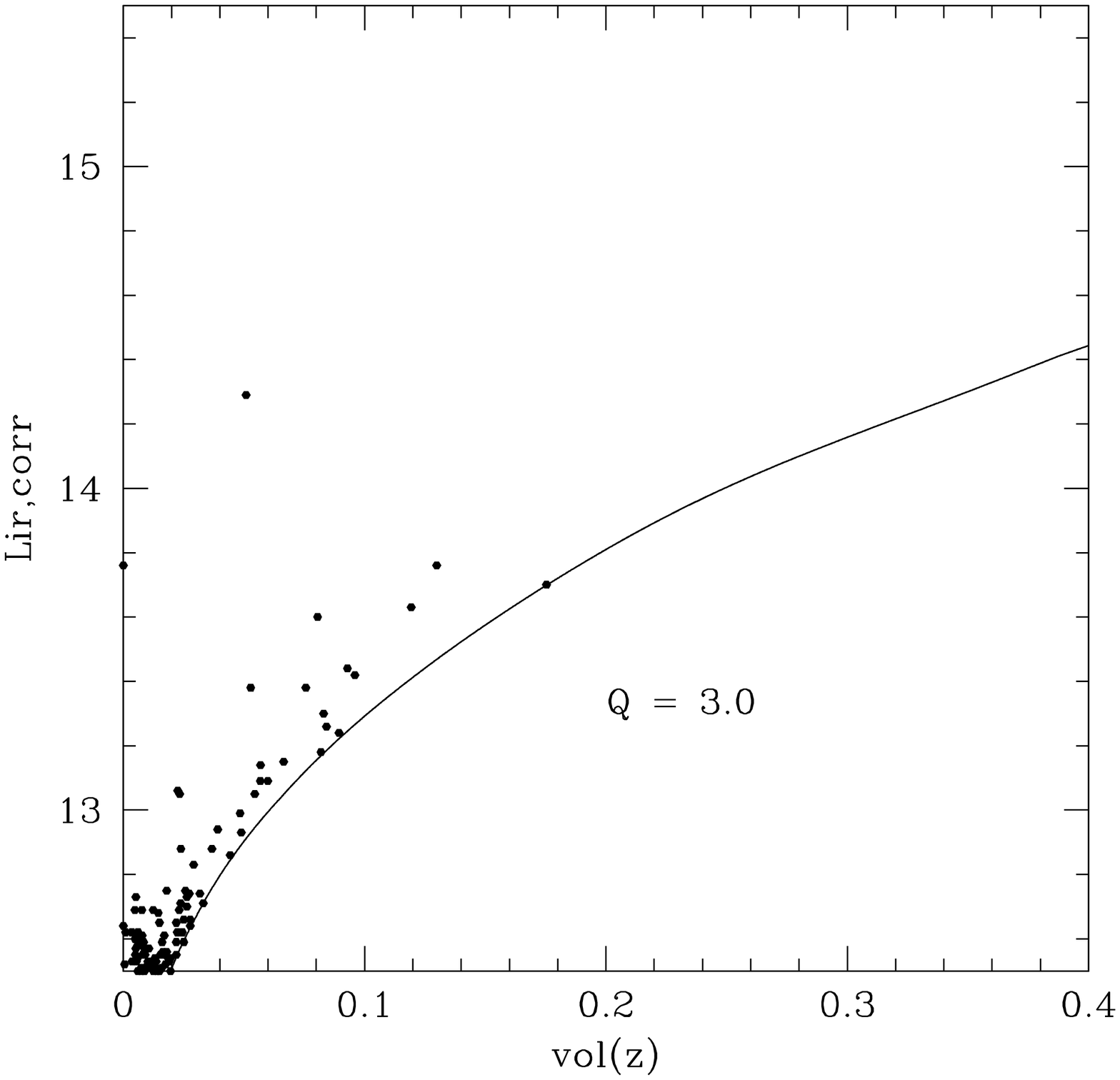,angle=0,width=7cm}
\caption{Infrared luminosity versus comoving volume for IIFSCz galaxies.  LH: no correction for evolution.
RH: corrected for luminosity evolution with Q=3.0.
}
\end{figure*}




Wang and Rowan-Robinson (2009b) presented luminosity functions for IIFSCz galaxies. Here we focus on the high luminosity 
end of the luminosity function using the $1/V_{\textrm{max}}$ method. For galaxies without an estimate of the best-fitting infrared 
template, we assume a SED of the form $S_{\nu}\propto \nu^{-2}$ to derive the K-correction term. Figure 9 shows the 60 $\mu$m 
luminosity function for IIFSSz galaxies with $L_{60} > 10^{12} L_{\odot}$ and $S60\geq 0.36$ Jy, divided into three redshift 
bins, 0.3-0.5, 0.5-1.0 and 1.0-2.0. The errors are derived from $\sigma = \sqrt{\Sigma_i \frac{weight,i}{V^2_{max,i}}}$, where each 
galaxy is weighted by the redshift completeness as a function of 60 micron flux density. Clearly the local 60 micron luminosity 
function (Wang \& Rowan-Robinson 2009b) indicated by the broken curve is not a good description of these ultraluminous and 
hyperluminous objects due to significant evolution. Figure 10 shows the corresponding luminosity function after correction for 
luminosity evolution of the form

$L(z) = L(0) \exp{(Q(t_0-t)/t_0)}$					(1)

with Q = 3.0. This is a higher rate of evolution than that deduced for the whole IIFSCz sample, Q = 1.7 (Wang \& Rowan-Robinson 2009b), 
for z $<$ 0.2.  A maximum-likelihood approach is also used to estimate the evolution of the luminosity function, following the procedure 
described in Wang \& Rowan-Robinson (2000b). We found a density evolution of the form $P=5.1$ or a luminosity evolution $Q=2.7$ 
in the redshift range $0.3<z<2.0$. For $L_{60} < 10^{12.5} L_{\odot}$ the luminosity function agrees well with the evolution-corrected 
luminosity function derived at lower luminosities, which is of the form derived by Saunders et al (1990). The evolution-corrected 
luminosity function then flattens towards higher luminosities, perhaps indicating the presence of a new population or a different 
physical mechanism is at work. Gravitational lensing is one such possible mechanism but we have argued above that rather 
few of the hyperluminous galaxies are in fact lensed. The very high star-formation rates in HLIRGs may be indicative of major 
merger events between massive galaxies. The analysis of star-formation rate versus gas mass deduced from CO observations 
by Rowan-Robinson (2000, Fig 20) may indicate a shorter time-scale for the starburst ($\sim 10^7$ yrs or less) than in lower 
luminosity starbursts.  The space-density of HLIRGs is $\sim 10^{-8} h_{72}^3 Mpc^{-3}$. 

In Wang and Rowan-Robinson (2009a), two methods were employed to estimate photometric redshifts. For sources with 2MASS 
J, H and Ks photometry, we choose redshifts given by the neural network training set method. For sources with SDSS ugriz 
photometry, redshifts derived from the template-fitting technique are preferred. In order to investigate the impact of photometric 
redshift errors, we carried out a Monte-Carlo analysis by generating random deviations from the estimated photometric redshifts 
using a Gaussian model with its standard deviation set to $4.4\%$ in $(1+z)$. The value $4.4\%$ is chosen because this is the 
worst rms error. We calculated the $1/V_{\textrm{max}}$ luminosity function of sources with $L_{60} > 10^{12} L_{\odot}$ and 
$S60\geq 0.36$ Jy for each of the one hundred Monte Carlo simulations created and found that the resulting rms error is several 
times smaller than the Poisson error. The outlier rate is estimated to be only  $<0.1\%$, but it might have an impact on our derived 
luminosity function, for example, in the highest redshift bin ($1.0<z<2.0$), where the total number of galaxies in the two luminosity 
bins is only three and four, respectively, of which three and two are photometric redshifts. 

\begin{figure*}
\epsfig{file=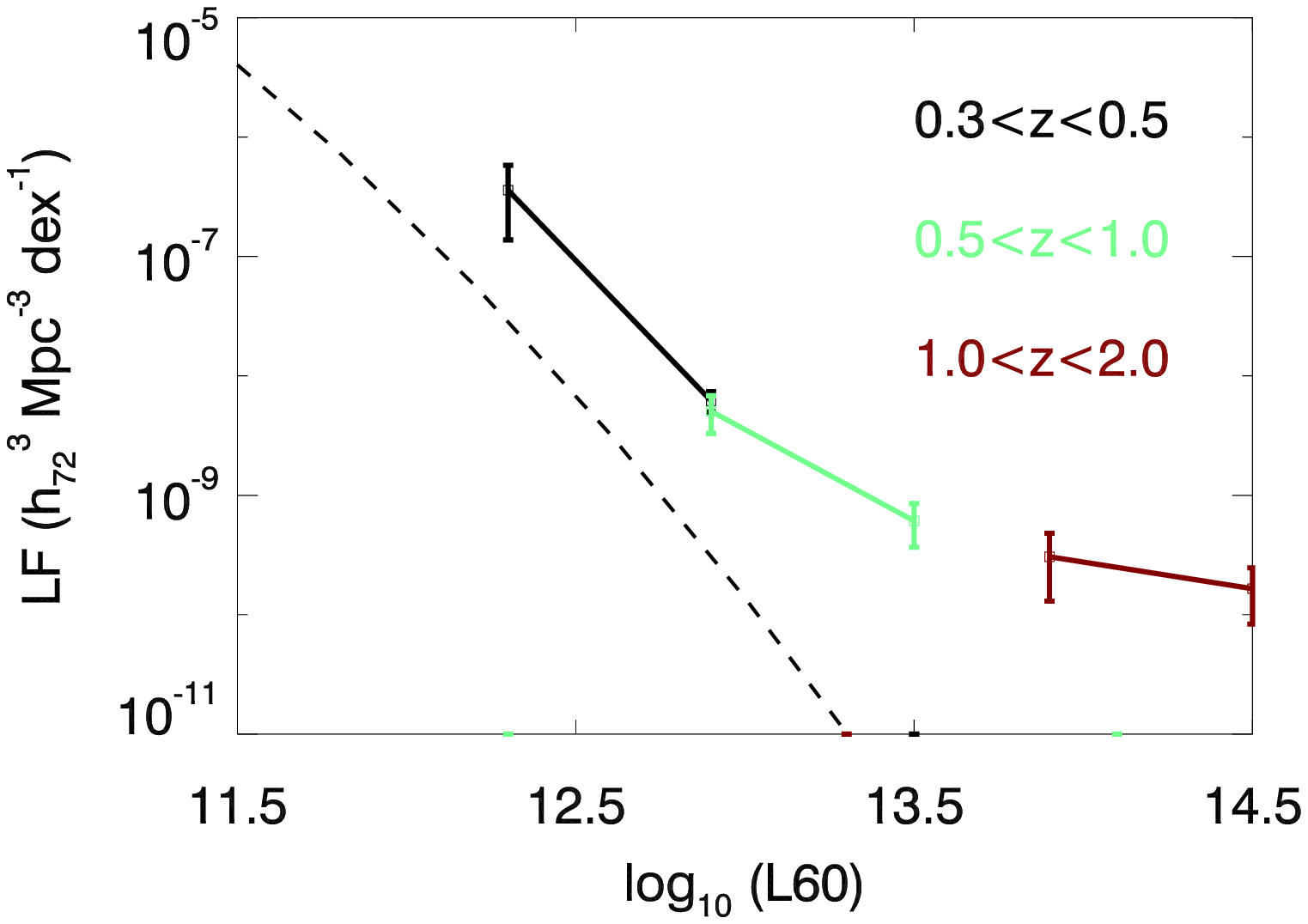,angle=0,width=12cm}
\epsfig{file=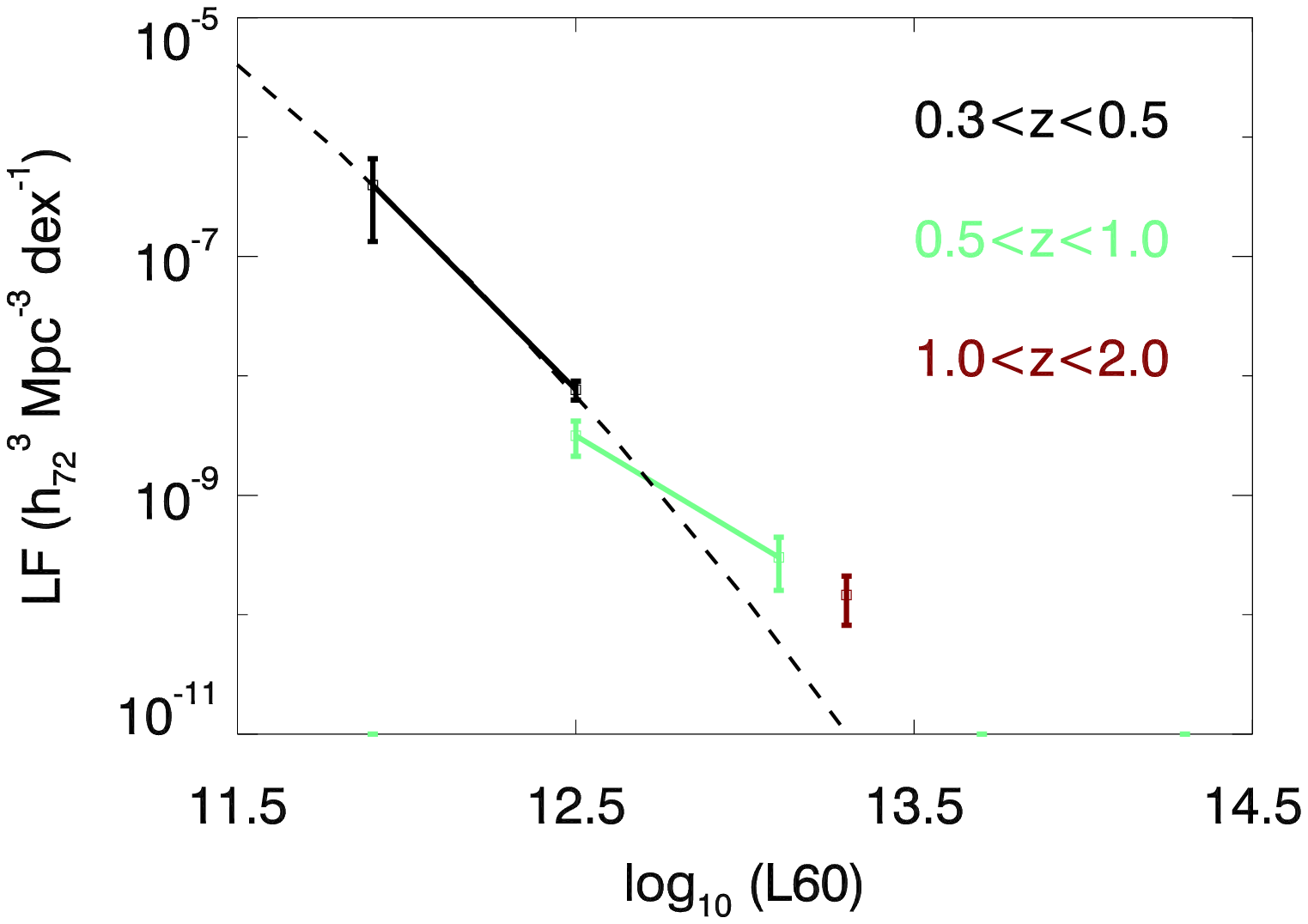,angle=0,width=12cm}
\caption{Upper panel : The 60 $\mu$m luminosity function for IIFSSz galaxies with $L_{60} > 10^{12} L_{\odot}$,
divided into three redshift bins, 0.3-0.5, 0.5-1.0 and 1.0-2.0. The broken curve is the luminosity function
calculated by Wang et al (2009) for z $<$ 0.2.
Lower panel: same, corrected for luminosity evolution of the form eqn (1) with Q = 3.0.
}
\end{figure*}


\section{Conclusions}
We present a catalogue of 179 hyperluminous infrared galaxies from the Imperial IRAS-FSS Redshift (IIFSCz)
Catalogue (Wang and Rowan-Robinson 2009a).  14 of these were included in the analysis of Rowan-Robinson 
(2000).   We use the infrared
template-fitting models of Wang and Rowan-Robinson (2009a) to predict fluxes at 350 and 850 $\mu$m.
Most would have 850 $\mu$m fluxes brighter than 5 mJy so should be readily detectable by current submillimetre
telescopes.  At least 15 $\%$ should be detectable in the {\it Planck} all-sky survey at 350 $\mu$m and 
all {\it Planck} all-sky survey sources with z $<$ 0.9 should be IIFSCz sources.

From the luminosity-volume test we find that HLIRGs show strong evolution.  A simple exponential luminosity 
evolution applied to all HLIRGs would be consistent with
the luminosity functions found in redshift bins 0.3-0.5, 0.5-1 and 1-2.  The evolution-corrected luminosity
function flattens towards higher luminosities perhaps indicating a different physical mechanism is at work
compared to lower luminosity starbursts.  In principle this could be gravitational lensing though previous 
searches with HST have not shown lensing to be widely prevalent in HLIRGs. 

Although for an interesting minority of HLIRGs the dominant contribution to the bolometric infrared luminosity
is due to an AGN dust torus, star formation appears to be the dominant factor in the overwhelming majority of cases,
with star-formation rates $>$ 1000 $M_{\odot} yr^{-1}$ being implied.  Even if the duration of these extreme
starbursts is shorter than that for less luminous starbursts (Rowan-Robinson 2000), they still represent an
extremely interesting and dramatic phase of galaxy evolution.

\section{Acknowledgements}
We thank Duncan Farrah for helpful discussions.  We thank the referee for helpful comments which led to 
significant improvements in this paper.


\end{document}